\documentclass[11pt,a4paper]{article}

\usepackage{cite}
\usepackage{amsmath, amsthm, amssymb,slashed}
\usepackage{ifpdf}
\ifpdf
  \usepackage[pdftex]{graphicx}
  \usepackage{epstopdf}
\else
  \usepackage[dvips]{graphicx}
\fi
\usepackage{fancyhdr}
\usepackage{mathbbol}
\usepackage{url}
\usepackage{color}
\usepackage{bbm}
\usepackage{dsfont}
\usepackage{bm}
\usepackage{amsfonts}

\def\be{\begin{equation}}
\def\ee{\end{equation}}
\def\bea{\begin{eqnarray}}
\def\eea{\end{eqnarray}}

\parskip=\baselineskip

\def\1{{\bf 1}}
\def\2{{\bf 2}}
\def\3{{\bf 3}}
\def\4{{\bf 4}}

\font\teneurm=eurm10 \font\seveneurm=eurm7 \font\fiveeurm=eurm5
\newfam\eurmfam
\textfont\eurmfam=\teneurm \scriptfont\eurmfam=\seveneurm
\scriptscriptfont\eurmfam=\fiveeurm

\font\teneusm=eusm10 \font\seveneusm=eusm7 \font\fiveeusm=eusm5
\newfam\eusmfam
\textfont\eusmfam=\teneusm \scriptfont\eusmfam=\seveneusm
\scriptscriptfont\eusmfam=\fiveeusm

\font\tencmmib=cmmib10 \skewchar\tencmmib='177
\font\sevencmmib=cmmib7 \skewchar\sevencmmib='177
\font\fivecmmib=cmmib5 \skewchar\fivecmmib='177
\newfam\cmmibfam
\textfont\cmmibfam=\tencmmib \scriptfont\cmmibfam=\sevencmmib
\scriptscriptfont\cmmibfam=\fivecmmib

\begin{document}
\begin{titlepage}
\begin{flushright}

\end{flushright}
\vskip 1.5in
\begin{center}
{\bf\Large{Hidden-Beauty Charged Tetraquarks and \\ Heavy Quark Spin Conservation}}
\vskip 0.5cm {A. Ali$^\P$, L. Maiani$^*$,  A.D. Polosa$^*$, V. Riquer$^*$}
 \vskip 0.05in {\textit{\small{$^\P$Deutsches Elektronen-Synchrotron DESY, D-22607 Hamburg, Germany}\vskip -.4cm  \textit{$^*$Dipartimento di Fisica, Sapienza Universit\`a di Roma, Piazzale Aldo Moro 2, I-00185 Roma, Italy}\vskip -.4cm
{\textit{and INFN Sezione di Roma, Piazzale Aldo Moro 2, I-00185 Roma, Italy }}}
}
\end{center}
\vskip 0.5in
\baselineskip 16pt
\begin{abstract}
Assuming the dominance of the spin-spin interaction
 in a diquark, we point out that the mass difference  in the beauty sector
 $M(Z_b^\prime)^\pm - M(Z_b)^\pm$ scales with quark masses as expected in QCD, with respect to the
corresponding mass difference $M(Z_c^\prime)^\pm - M(Z_c)^\pm$.
Notably, we show that the decays $\Upsilon(10890) \to \Upsilon(nS) \pi^+\pi^-$ and  $\Upsilon(10890) \to (h_b(1P), h_b(2P)) \pi^+\pi^-$ are compatible with
heavy-quark spin conservation if the contributions of $Z_b,Z_b^\prime$ intermediate states are taken
into account, $\Upsilon(10890)$ being  either a $\Upsilon(5S)$ or the beauty analog of $Y_c(4260)$.  Belle results on these decays support the quark spin wave-function of the Z states as tetraquarks. 
We also consider the role of 
light quark spin non-conservaton in $Z_b,Z_b^\prime$ decays into $B B^*$ and $B^* B^*$. 
 Indications of possible signatures of the still missing $X_b$ resonance are proposed.
\newline
\newline
Preprint \# DESY 14-234
\end{abstract}
%
\end{titlepage}

Tetraquark interpretation of the hidden charm and beauty exotic resonances has been advanced and studied
in considerable detail (see Refs.~\cite{noi05}~\cite{brodsky}, and~\cite{ali}).
In a recent contribution~\cite{noi14},  a new scheme for the spin-spin quark interactions in
the  hidden charm resonances has been proposed, which reproduces well the 
 mass and decay pattern of   $X(3872)$, of the recently discovered~\cite{bes3} $Z_c^{\pm,0}(3900)$, $Z_c^{\pm,0}(4020)$,  and of the lowest lying $J^{PC}=1^{--}$ $Y$ states. 

Tetraquark states in the large $N_c$ (color) limit of QCD have been considered in~\cite{weinberg} and~\cite{peris} (see also the review~\cite{review} and references therein).
Compact tetraquark mesons may have decay widths as narrow as $1/N_c$, contrary to previous beliefs,
 and therefore they are
reasonable candidates for a secondary spectroscopic meson series, in addition to the standard $q\bar q$ one. Another route to multiquark meson states is being explored in ~\cite{braaten} within the Born-Oppenheimer approximation of QCD , where examples of selection rules for hadronic transitions have been worked out. 

In this letter we consider the extension of the scheme presented  in~\cite{noi14}
for the hidden-charm to  the hidden-beauty resonances $Z_b^{\pm,0}(10610)=Z_b$ and $Z_b^{\pm,0}(10650)=Z_b^\prime$.

These resonances are interpreted as $S-$wave $J^{PG}=1^{++}$ states with diquark spin distribution (use the notation $|s_{[bq]},s_{[\bar b\bar q]}\rangle$ for diquark spins)
\bea
&&|Z_b\rangle=\frac{|1_{bq},0_{\bar b\bar q}\rangle-|0_{bq},1_{\bar b\bar q}\rangle}{\sqrt{2}}\notag \\
&&|Z_b^\prime\rangle=|1_{b q},1_{\bar b\bar q}\rangle_{J=1}\label{prima}
\eea
The $J^P=1^+$ multiplet is completed by $X_b$, which is given by the $C=+1$ combination
\be
|X_b\rangle=\frac{|1_{bq},0_{\bar b\bar q}\rangle+|0_{bq},1_{\bar b\bar q}\rangle}{\sqrt{2}}
\ee
Assuming  the spin-spin interaction inside diquarks to dominate, we expect $X_b$ and $Z_b$ to be degenerate, with 
$Z_b^\prime$ heavier according to 
\be
M(Z_b^\prime)-M(Z_b)=2\kappa_b
\ee
where $\kappa_b$  is the strength of the spin-spin interaction inside the diquark. A similar analysis
 for the hidden-charm resonances has produced the value~\cite{noi14}
\be
2\kappa_c=M(Z_c^\prime)-M(Z_c)\simeq 120~{\rm MeV}
\ee
The QCD expectation is $\kappa_b:\kappa_c=M_c:M_b$. The ratio can be estimated from the masses reported in~\cite{pdg}
\be
\frac{M_c}{M_b}\simeq \frac{1.27}{4.18}=0.30
\ee
giving $2 \kappa_b \simeq 36$~MeV, which fits nicely with the
observed  $Z_b^\prime-Z_b$ mass difference ($\simeq 45$ MeV).
 
Next we consider another crucial prediction of QCD, namely conservation of the heavy quark spin in hadronic decays. 
 
 We recall that $Z_b,Z_b^\prime$ are observed in the decays of $\Upsilon(10890)$
 \be
\Upsilon(10890)\to Z_b/Z_b^\prime+\pi\to h_b(nP)\pi\pi \label{eq:udecay} 
 \ee 
The $\Upsilon(10890)$ is usually reported as the $\Upsilon(5S)$ since its mass is close to the mass of the $5S$ state predicted by potential models. However,
a different assignment was proposed in~\cite{aliZbs}, namely $\Upsilon(10890)=Y_b$, the latter state
 being a $P-$wave tetraquark 
analogous to the $Y(4260)$.
A reason for this assignment is the analogy of $\Upsilon(10890)$ decay (\ref{eq:udecay}) with  $Y(4260)\to Z_c(3900)+\pi$,
with  $Y(4260)$ being the the first discovered $Y$ state~\cite{babary}.
Current experimental situation about $\Upsilon(10890)$ is still in a state of flux. In our opinion,
the possibility that $\Upsilon(10890)$ is an unresolved peak involving both the $\Upsilon(5S)$ and
$Y_b$, reported by Belle some time ago~\cite{Chen:2008xia}, is plausible,
 providing a  resolution of the observed branching ratios measured at the
 $\Upsilon(10890)$~\cite{tamponi}. However, this identification is not a
requirement in the considerations presented below.
In fact, following the assignment of 
$Y(4260)$ as  a $P-$wave tetraquark with $s_{c\bar c}=1$~\cite{noi14}, one sees that in both cases the initial state in~(\ref{eq:udecay}) corresponds to 
$s_{b\bar b}=1$. As is well known $h_b(nP)$ has $s_{b\bar b}=0$, pointing to a possible violation of
the heavy-quark spin conservation, as suggested in~\cite{tamponi}.

We show now that the contradiction is only apparent.
Expressing the states in the the basis of definite $b\bar b$ and $q\bar q$ spin, one finds
\bea
&&|Z_b\rangle=\frac{|1_{q\bar q},0_{b\bar b}\rangle-|0_{q\bar q},1_{b\bar b}\rangle}{\sqrt{2}}\notag\\
&&|Z_b^\prime\rangle=\frac{|1_{q\bar q},0_{b\bar b}\rangle+|0_{q\bar q},1_{b\bar b}\rangle}{\sqrt{2}}
\label{eq:scomposition}
\eea
It is conceivable that the subdominant spin-spin interactions may play a non negligible role
 in the $b$-systems, as the spin-spin dominant interaction is suppressed by the large $b$ quark mass.
 In this case the composition of $Z_b,Z_b^\prime$ indicated in  Eq.~(\ref{eq:scomposition}) would be more general:
 \bea
&&|Z_b\rangle=\frac{\alpha|1_{q\bar q},0_{b\bar b}\rangle-\beta|0_{q\bar q},1_{b\bar b}\rangle}{\sqrt{2}}\notag\\
&&|Z_b^\prime\rangle=\frac{\beta|1_{q\bar q},0_{b\bar b}\rangle + \alpha |0_{q\bar q},1_{b\bar b}\rangle}{\sqrt{2}}
\label{eq:scompositiongen}
\eea

The kets in (\ref{eq:scomposition}) and (\ref{eq:scompositiongen}) represent a superposition of color singlet and color octet. These states evolve, by QCD interactions, into pairs of color singlet mesons with heavy quark spin conserved in the limit $M_b\to \infty$, e.g. $|1_{q\bar q},0_{b\bar b}\rangle$ gives rise to $\eta_b, h_b$ etc., but not $\Upsilon_b, \chi_{jb}$ etc.. No similar constraint applies to the light quark spin.

We assume $\alpha$ and $\beta$ to be both different from zero. If either one of the two vanishes, the decay~(\ref{eq:udecay}) would be altogether forbidden by heavy quark spin conservation, contrary to what
is observed for the distribution of $M(h_b\pi)$ in Ref.~\cite{belle5S}.

Define
\bea
&&g_{Z}=g(\Upsilon \to Z_b\pi)g(Z_b\to h_b\pi)\propto -\alpha\beta \langle h_b|1_{q\bar q},0_{b\bar b}\rangle \langle 0_{q\bar q},1_{b\bar b}|\Upsilon\rangle\notag \\ 
&&g_{Z^\prime}=g(\Upsilon \to Z_b^\prime\pi)g(Z_b^\prime \to h_b\pi)\propto \alpha\beta \langle h_b|1_{q\bar q},0_{b\bar b}\rangle \langle 0_{q\bar q},1_{b\bar b}|\Upsilon\rangle
\label{eq:ampl}
\eea
where $g$ are the effective strong couplings at the vertices
 $\Upsilon\, Z_b\, \pi$ and $Z_b\, h_b\, \pi$.
For both assignments of $\Upsilon(10890)$, Eq.~(\ref{eq:scompositiongen}) and heavy quark
 spin conservation require
\be
g_Z=-g_{Z^\prime}
\ee 
independently from the values of the mixing coefficients $\alpha,\beta$. 

In Ref.~\cite{belle5S} the amplitude for the decay~(\ref{eq:udecay}) is fitted with two Breit-Wigners corresponding to the $Z_b,Z_b^\prime$ intermediate states.
Table~I therein, that we transcribe here in Table~1, shows the relative normalizations
 and phases obtained by the fit, for decays into $h_b(1P)$ and $h_b(2P)$.
\begin{table}
\begin{center}
\begin{tabular}{| l | c | c | c | c | c | }
\hline
    & & & & &  \\
   Final State & $\Upsilon(1S)\pi^+\pi^-$ & $\Upsilon(2S)\pi^+\pi^-$ & $\Upsilon(3S)\pi^+\pi^-$ &  $h_b(1P)\pi^+\pi^-$ & $h_b(2P)\pi^+\pi^-$  \\
    & & & & &  \\
   \hline\hline
    & & & & &  \\
  Rel. Norm. & $0.57\pm 0.21^{+0.19}_{-0.04}$ & $0.86\pm 0.11^{+0.04}_{-0.10}$ & $0.96\pm 0.14^{+0.08}_{-0.05}$ & $1.39\pm 0.37^{+0.05}_{-0.15}$ & $1.6^{+0.6+0.4}_{-0.4-0.6} $ \\
   & & & & &  \\
  \hline
    & & & & & \\
  Rel. Phase & $58\pm43^{+4}_{-9}$ & $-13\pm13^{+17}_{-8}$ & $-9\pm 19^{+11}_{-26}$ &  $187^{+44+3}_{-57-12}$& $181^{+65+74}_{-105-109}$ \\
 & & & & &  \\
  \hline
\end{tabular}
  \end{center} \caption{ Values of the relative normalizations and of the relative phases (in degrees), for $s_{b\bar b}:1\to 1$ and $1\to 0$ transitions, as reported by~\cite{belle5S}.\label{eq:tav}}
\end{table}
Within large errors, consistency with Eq.~(\ref{eq:ampl}), that is with the heavy-quark spin
 conservation, is apparent.

It is interesting that the same conclusion was drawn using a picture in which $Z_{b},Z_{b^\prime}$ have a 
 ``molecular''  type structure~ \cite{voloshin} 
\be
Z_b=\frac{|B , \bar B^*\rangle -|\bar B , B^*\rangle }{\sqrt{2}}\quad\quad Z^{\prime}_b=|B^*,\bar B^* \rangle_{J=1}
\ee

To determine $\alpha$ and $\beta$ separately, one has to resort to $s_{b \bar b} : 1 \to 1$ transitions,
 such as
\be
 \Upsilon(10890)\to  Z_b/Z_b^\prime+\pi\to  \Upsilon(nS)\pi\pi~(n=1,2,3)
 \label{1to1}
 \ee
The   effective couplings  analogous to~(\ref{eq:ampl}) are
\bea
&&f_{Z}=f(\Upsilon \to Z_b\pi)f(Z_b\to \Upsilon(nS)\pi)\propto |\beta|^2 \langle \Upsilon(nS)|0_{q\bar q},1_{b\bar b}\rangle \langle 0_{q\bar q},1_{b\bar b}|\Upsilon\rangle\notag \\ 
&&f_{Z^\prime}=f(\Upsilon \to Z_b^\prime\pi)f(Z_b^\prime\to \Upsilon(nS)\pi)\propto |\alpha|^2 \langle \Upsilon(nS)|0_{q\bar q},1_{b\bar b}\rangle \langle 0_{q\bar q},1_{b\bar b} |\Upsilon\rangle\notag 
\label{eq:ampl2}
\eea
The Dalitz plot of these decays indicate indeed that the transitions~(\ref{1to1}) proceed mainly
 through $Z_b$ and $Z_b^\prime$~\cite{belle5S,tamponi}, though the amplitude for the process
 $\Upsilon(10890) \to \Upsilon(1S)\pi^+\pi^-$ has a significant direct component, which is
expected in the tetraquark interpretation of the state $\Upsilon(10890)$~\cite{ali}. This is also
reflected in Table~1. A quantitative analysis of the Belle data
including the direct and resonant components (i.e., via the intermediate resonant states $Z_b$
and $Z_b^\prime$) is required to test the underlying dynamics. Leaving this for the future, 
we argue here that parametrizing the amplitude in terms of two Breit-Wigner as before, 
 one determines the ratio $\alpha/\beta$.
 Indeed, from the Belle results~\cite{belle5S} we find the following weighted average values\footnote{for simplicity we have computed the weighted averages with statistical errors only.}:  
\bea 
&&s_{b\bar b}:1\to 1~{\rm transition}:\notag \\
&&\overline{{\rm Rel. Norm.}}= 0.85\pm0.08=|\alpha|^2/|\beta|^2\notag \\
&&\overline{{\rm Rel. Phase}}= (-8\pm10)^\circ
\eea
and   
\bea 
&&s_{b\bar b}:1\to 0~{\rm transition}:\notag \\
&&\overline{{\rm Rel. Norm.}}= 1.4\pm0.3\notag \\
&&\overline{{\rm Rel. Phase}}= (185\pm42)^\circ.
\eea
Within errors, the tetraquark assignment in Eqs.~(\ref{prima}) and~(\ref{eq:scomposition}) with $\alpha=\beta=1$ is supported.

As a side remark, we observe that a Fierz rearrangement  similar to the one used in~(\ref{eq:scomposition}) puts together $b\bar q$ and $q\bar b$ fields
\bea 
&&|Z_b\rangle = |1_{b\bar q}, 1_{q \bar b}\rangle_{J=1} \notag\\
&&|Z_b^\prime\rangle=\frac{|1_{b\bar q}, 0_{q \bar b}\rangle + |0_{b\bar q}, 1_{q \bar b}\rangle}{\sqrt{2}} \label{ultima}
\eea
The labels $0_{b\bar q}$ and $1_{b \bar q}$ could be viewed as indicating
 $B$ and $B^*$ mesons, respectively, leading to
the prediction of the decay patterns $Z_b\to B^{*}\bar{B}^{*}$ and  $Z_b^\prime\to B\bar{B}^{*}$~\cite{ali}.
This would not be in agreement with the Belle data~\cite{tamponi}. 

This argument, however, rests  on conservation of the light quark spin which, unlike the heavy quark spin, may change when the color octet pairs which appear in~(\ref{ultima}), evolve 
into pairs of color singlet mesons.  Therefore predictions derived from~(\ref{ultima}) are not as reliable as those derived from~(\ref{eq:scomposition}). We also stress that the issue of the unaccounted 
direct production of the $B^{*}\bar{B}^{*}$, $ B\bar{B}^{*} $ and related states in the Belle data,
collected at and near the $\Upsilon(10890)$ resonance~\cite{tamponi}, once satisfactorily resolved,
may also reflect on the resonant contributions  $Z_b\to B^{*}\bar{B}^{*}$ and  $Z_b^\prime\to B\bar{B}^{*}$.
We look forward to an improved analysis of the Belle measurements.

Finally we comment on the expected positive charge conjugation  state, $X_b$. On the basis of the assumed spin-spin interaction, one predicts $M(X_b)\simeq M(Z_b)\simeq 10600$~MeV. Such a state 
has been searched by CMS~\cite{cmsb} in the region $10.1<M<11.0$~GeV  and by ATLAS~\cite{atlasb} in the region $10.5<M<11.0$~GeV  looking for the decay
\be
X_b\to \Upsilon(1S)\pi\pi
\ee
so far with negative results. 

In Ref.~\cite{brodsky}, it is noted that the near equality of the branching ratios for $X(3872)\to J/\psi \,2\pi$ and $X(3872)\to J/\psi\, 3\pi$ can be understood 
if  $X(3872)$ is predominantly isosinglet. The isospin allowed decay in $J/\psi \,\omega$ is phase space forbidden and the decay in the $J/\psi \,\rho$ mode, although isospin forbidden, is phase space favoured, leading to similar rates.

In the $X_b$ decay, both $\omega$ and $\rho$ channels are allowed by phase space, so that, if $X_b$ is isosinglet, the dominant mode would be into $\Upsilon(1S) \, \omega$. The suggestion therefore is to look at the
 decay $X_b(10600) \to \Upsilon(1S) \, 3\pi$ with the $3\pi$ in the $\omega$ mass band, in parallel with the search for the $X_b(10600) \to \Upsilon(1S)\, 2\pi$ channel with the $2\pi$ in the $\rho$ band.  A search for 
 $Y(10890)\to \gamma X_b\to \gamma\omega \Upsilon(1S)$ has been presented in Ref.~\cite{belleb}  leading for the moment to an upper bound only.

\section*{Acknowledgements}
We thank C. Hanhart  for interesting discussions.

\bibliographystyle{unsrt}

\end{document}